\documentclass[conference]{IEEEtran}
\IEEEoverridecommandlockouts
\usepackage{algorithm}
\usepackage{algorithmic}
\usepackage{multirow}
\usepackage{hyperref}

\usepackage{graphicx}
\usepackage{epstopdf}
\usepackage{subcaption}
\usepackage{textcomp}
\usepackage{xcolor}
\usepackage{soul}
\usepackage{float}

\usepackage{tikz}

\AtBeginDocument{%
  \providecommand\BibTeX{{%
    \normalfont B\kern-0.5em{\scshape i\kern-0.25em b}\kern-0.8em\TeX}}}
    
\let\oldnl\nl
\newcommand\nonl{
  \renewcommand{\nl}{\let\nl\oldnl}}

\newcommand*\circled[1]{\tikz[baseline=(char.base)]{ \node[shape=circle,fill,inner sep=0.4pt] (char) {\textcolor{white}{#1}};}}

\usepackage{fancyhdr}

\begin{document}

\pagestyle{fancy}
\fancyhead[C]{This work has been accepted and presented on IEEE International Conference on High Performance Computing 2022 (HiPC)}

\title{Energy Consumption Evaluation of Optane DC Persistent Memory for Indexing Data Structures}


\author{\IEEEauthorblockA{Manolis Katsaragakis$^{\star \dagger}$, Christos Baloukas$^{\star}$, Lazaros Papadopoulos$^{\star}$, Verena Kantere$^{\bullet}$, \\Francky Catthoor$^{\dagger \circ}$, Dimitrios Soudris$^{\star}$}
\IEEEauthorblockA{$^{\star}$\textit{Microprocessors and Digital Systems Laboratory, ECE , National Technical University of Athens, Greece}\\
$^{\dagger}$\textit{Katholieke Universiteit Leuven, Kasteelpark Arenberg 10, 3001 Heverlee, Belgium}\\
$^{\bullet}$\textit{Knowledge and Database Systems Laboratory, ECE , National Technical University of Athens, Greece}\\
$^{\circ}$\textit{IMEC, Kapeldreef 75, 3001 Heverlee, Belgium}\\
$^{\star}$\{mkatsaragakis, lpapadop, cmpalouk,  dsoudris\}@microlab.ntua.gr}
$^{\bullet}$vkante@dblab.ece.ntua.gr, $^{\dagger}${francky.catthoor@esat.kuleuven.be}

\thanks{This work has been partially funded by EU Horizon 2020 program under grant agreement No 101015922 AI@EDGE (https://aiatedge.eu/).}}

\maketitle

\maketitle

\begin{abstract}
The Intel Optane DC Persistent Memory (DCPM) is an attractive novel technology for building storage systems for data intensive HPC applications, as it provides lower cost per byte, low standby power and larger capacities than DRAM, with comparable latency. 
This work provides an in-depth evaluation of the energy consumption of the Optane DCPM, using well-established indexes specifically designed to address the challenges and constraints of the persistent memories. 
We study the energy efficiency of the Optane DCPM for several indexing data structures and for the LevelDB key-value store, under different types of YCSB workloads. By integrating an Optane DCPM in a memory system, the energy drops by 71.2\% and the throughput increases by 37.3\% for the LevelDB experiments, compared to a typical SSD storage solution.
\end{abstract}


\section{Introduction} 
\label{sec:introduction}

The memory system is one of the main components that limit the scalability and contribute to the energy consumption of supercomputers~\cite{patil2019performance}.  The integration of more DRAM modules to enable more complex simulations, analytics and effective in-memory processing has negative impact on the sustainability and maintenance costs of supercomputing centres. In particular, despite the low access latency of traditional DRAM technologies, the increased leakage and refresh power requirements limit DRAM scalability and introduce a significant challenge towards reaching exascale performance. 

In order to overcome DRAM limitations, non-volatile memory (NVM) technologies have been introduced, such as the 3D-XPoint, which is a subclass of the Phase-Change Memories (PCM)~\cite{wong2010phase}, Spin-Transfer Torque RAM (STT-RAM)~\cite{kultursay2013evaluating} and Resistive RAM (ReRAM)~\cite{sheu2010fast}. State-of-the-Art commercial platforms integrate Intel Optane DC Persistent Memory (DCPM) modules along with DRAM, leading to heterogeneous memory systems~\cite{izraelevitz2019basic}. For instance, the upcoming Aurora exascale supercomputer employs the DAOS storage architecture, which integrates a complex memory and storage hierarchy, including Intel Optane DCPM modules~\cite{daos}. 

These emerging memory technologies provide higher density than DRAM, enabling increased aggregate memory capacities with fewer nodes, having positive impact on the energy consumption, resilience and sustainability. Additionally, the data persistence features of the NVM technologies can be used to provide fault tolerance support to applications. On the other hand, the Optane DCPM provides, in general, higher access latency and lower bandwidth compared to DRAM~\cite{izraelevitz2019basic, weiland2019early, xiang2022characterizing}. However, recent studies indicate that the performance of the Optane DCPM depends a lot on the workload access pattern and size~\cite{xiang2022characterizing}. 

The contribution of the Optane DCPM technology to enable more complex and effective HPC simulations has been investigated in several recent works ~\cite{weiland2019early, patil2019performance, mironov2019performance, peng2020demystifying, venkatesh2021scheduling}. They indicate that using Optane DCPM as a volatile memory and DRAM as a cache enables close to DRAM performance. However, directly replacing the DRAM with Optane DCPM, without the use of DRAM as a cache, significantly reduces performance, due to effects such as the write throttling and concurrency contention~\cite{peng2019system}. 

Alternatively, the Optane DCPM can be used as a persistent storage medium to enable the storage and processing of massive volumes of scientific data gathered by simulations or instruments~\cite{smart2019high, ejarque2022enabling}. 
%
Recent studies indicate that in order to take advantage of the scalability opportunities that storage systems enabled by persistent memories can offer to applications, the indexing data structures of the storage system need to be adapted in order to address the challenges and limitations that the NVM technologies impose~\cite{levandoski2013bwtree, oukid2016fptree, chen2015wbtree}. For example, a fundamental challenge when NVMs are employed for persistent storage is the data consistency. Since modern CPUs are designed for volatile DRAM architectures, they typically cache and reorder memory writes, as they target higher performance. Therefore, existing indexing structures, designed for DRAM only systems, cannot be directly deployed on heterogeneous hierarchies, as they do not take into account the persistent nature of NVMs, as well as other challenges, such as the asymmetry in terms of read and write latency and the limited write endurance of NVMs. 

Towards this direction, several works exist in the literature that propose indexes specifically designed for NVMs~\cite{levandoski2013bwtree, oukid2016fptree, chen2015wbtree, hwang2018fastfair, mao2012masstree}. Some recent works provide performance-based evaluation of B+ tree indexing data structures for persistent memories~\cite{lersch2019evaluating,he2022evaluating}, while others convert DRAM-based B+ trees, tries, radix trees, and hash table indexes into NVM-oriented and evaluate their impact in terms of metrics such as the throughput and cache utilization~\cite{lee2019recipe}. 

This work targets HPC developers who consider the use of Optane DCPM as a persistent storage for applications which access large amounts of data through indexing data structures. 
The existing works in the literature, either focus on the evaluation of the Optane DCPM as a volatile main memory, or provide a performance characterization of its persistence features using the Optane DCPM as a storage device. 
Although these works are very relevant, they still miss the important energy component in the global analysis. Therefore, to address this gap, this work focuses on the energy consumption evaluation of the Optane DCPM configured as a persistent storage. 
The two major contributions of this work are the following: 
\begin{itemize}
    \item A thorough evaluation of the Optane DCPM as a storage medium, in terms of energy consumption, using representative and well-established B+ tree indexing data structures, triggered by various types of YCSB workloads. 
    \item A combined performance and energy consumption evaluation of the LevelDB key-value store deployed on an Optane DCPM and an energy and throughput comparison with the corresponding deployment on an SSD.
\end{itemize}

To the best of our knowledge, this is the first work that entirely focuses on the energy efficiency aspects of the Optane DCPM. Although there exists several works that investigate the performance behavior of this new memory architecture, as detailed in the following sections, the energy consumption characteristics of the Optane DCPM have been evaluated to a very limited extent, yet.

The rest of this paper is organized as follows:
Section \ref{sec:related_work} discusses the related work, focusing on the main differentiations  between the present and the existing works. Section \ref{sec:optane-overview} is an overview of the Optane DCPM technology. The evaluation methodology is described in Section \ref{sec:evaluation_methodology}. Section \ref{sec:evaluation} presents a set of baseline scalability results and the energy consumption analysis based on B+ tree indexing data structures and on the LevelDB key-value store. Additionally, it highlights the main observations and open research directions. Finally, in Section \ref{sec:conclusion} we draw conclusions. 
\section{Related Work} 
\label{sec:related_work}

The existing works related to the Optane DCPM can be loosely classified into three categories: 
\begin{itemize}
    \item Empirical evaluation of the Optane DCPM performance through microbenchmarks and/or applications and investigation of challenges and limitations~\cite{izraelevitz2019basic, weiland2019early, patil2019performance, mironov2019performance, peng2019system, peng2020demystifying, lee2019recipe, lersch2019evaluating, he2022evaluating}
    \item New applications and data structures optimized for the Optane DCPM as persistent storage (mainly novel tree-based data structures)~\cite{zhou2019dptree, chen2020utree}
    \item Novel workload schedulers, data placement algorithms and other runtimes for the Optane DCPM~\cite{weiland2021usage, xiang2022characterizing, katsaragakis2022memory}
\end{itemize}

The present work is mostly related to first category, as it aims to thoroughly evaluate Optane DCPM using indexing data structures. However, in contrast to the existing works, it focuses on the energy consumption aspects. 

Recently, the Optane DCPM was extensively evaluated using real-world scientific applications. The typical hardware configuration applied in these works was the use of the Optane DCPM as the main memory, while DRAM was used as a cache. The persistent features of the Optane DCPM are not available in this case. Recent works conclude that the performance gap between an Optane DCPM and a DRAM-only system is relatively small for this hardware configuration~\cite{peng2019system, mironov2019performance}. Works that entirely replace DRAM with Optane DCPM (i.e. DRAM was entirely deactivated) highlight the fact that there is a significant performance overhead, due to the Optane DCPM bandwidth saturation for real-world multi-threaded applications~\cite{izraelevitz2019basic, patil2019performance}.

In contrast to the aforementioned works, using Optane DCPM as a persistent memory requires changes in the application source code. Additionally, the API for placing data on an Optane DCPM is available for dynamically allocated data structures only (e.g. through \textit{libvmmalloc})~\cite{weiland2019early}. Therefore, this configuration is only partially evaluated in the existing literature using real-world scientific applications~\cite{peng2020demystifying, peng2019system}. The majority of works evaluate the Optane DCPM in this configuration either using microbenchmarks, such as the Stream Triad or mini-applications~\cite{weiland2019early}. A more extensive evaluation, considering both the performance and energy consumption is available in the literature, but targets graph applications specifically~\cite{peng2019system}. On the other hand, there are several works which evaluate B+ tree indexing data structures on the Optane DCPM, as a persistent memory. They typically compare indexes in terms of application-level performance-related metrics, such as throughput and latency~\cite{lee2019recipe, lersch2019evaluating, he2022evaluating}. 

To summarize, in contrast with the existing studies, the present work evaluates the Optane DCPM as a persistent memory, in terms of energy consumption using B+ tree indexes and also provides a combined performance and energy consumption analysis of the LevelDB key-value store.
\section{Overview of Optane DCPM} 
\label{sec:optane-overview}

The Intel Optane DCPM is the first commercially available NVDIMM. It is based on the 3D-XPoint technology and provides both byte-addressability and persistence. From CPU perspective, the access granularity is 64 bytes. However, the physical media access granularity is 256 bytes. Therefore, write accesses smaller than 256 bytes are read-modify-write operations, which results in write amplification. The effects of write amplification have been extensively studied in the literature~\cite{peng2020demystifying}. 

\begin{figure}[t]
       \begin{minipage}[]{\linewidth}
       \centering   
        \begin{subfigure}[]{0.43\columnwidth}
                \includegraphics[width=\textwidth]{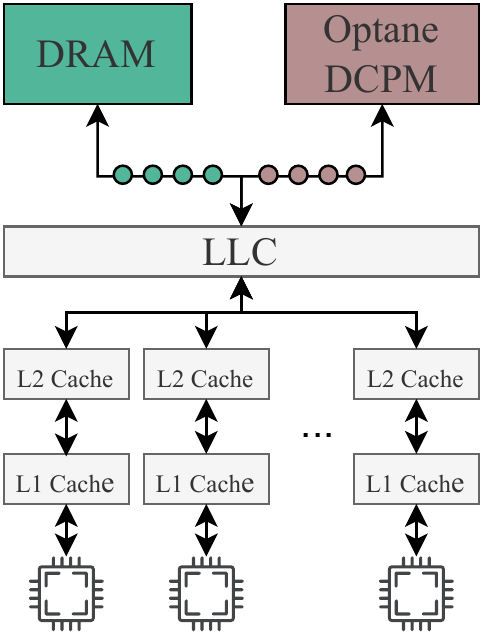}
                \caption{AppDirect Mode}
                \label{fig:app-direct}
        \end{subfigure}
        ~
        \begin{subfigure}[]{0.43\linewidth}
                \includegraphics[width=\textwidth]{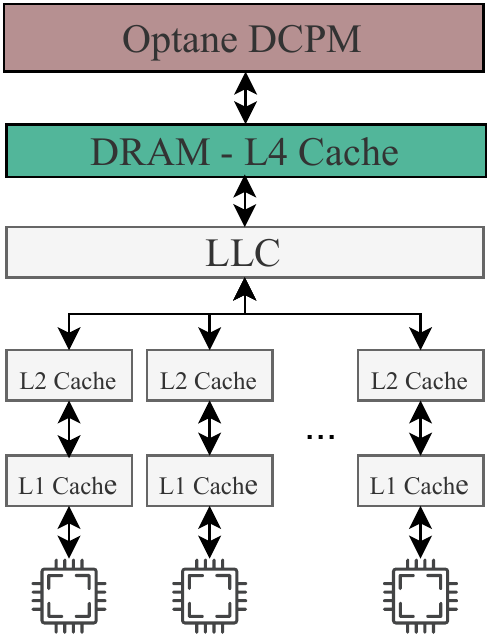}
                \caption{Memory Mode}
                \label{fig:memory-mode}
        \end{subfigure}
        \caption{Overview of Intel Optane DCPM operating modes}
        \label{fig:optane-arch}
        \end{minipage} 
\end{figure}

The Optane DCPM provides much higher capacity and lower power than DRAM, but, in general, at cost of increased latency and lower bandwidth. Fig.~\ref{fig:optane-arch} depicts an heterogeneous memory system for the alternative operating modes of the Optane DCPM. The Optane DCPM can be configured to support either the \textit{AppDirect} (Fig.~\ref{fig:app-direct}) or the \textit{Memory mode} (Fig.~\ref{fig:memory-mode}). In the AppDirect mode, Optane DCPM operates as a persistent storage. Access to the storage space is similar to memory mapped file operations. In \textit{Memory Mode}, Optane DCPM operates as the main memory space without persistent properties. In this mode, the DRAM is used as an extra layer of cache on top of the Last-Level Cache. 

In the \textit{AppDirect} mode, developers can rely on the Persistent Memory Development Kit (PMDK) for persistent memory allocations~\cite{pmdk}. Several libraries such as the \textit{libvmmalloc} and \textit{pmemobj} are provided that facilitate the dynamic allocation of objects that are traditionally handled by \textit{malloc}, \textit{free} and \textit{memalign}. For instance, \textit{libvmmalloc} allows a transparent runtime mapping of all dynamic allocations to the persistent memory. Thus, dynamically allocated data structures are placed into a memory mapped file, which is stored in the persistent memory. However, in \textit{Memory mode}, all data objects are allocated to the Optane DCPM by default, transparently to developers. In this work, the Optane DCPM is configured in the \textit{AppDirect} mode, as we aim at using it as a persistent storage for data intensive applications, instead of a volatile memory.

\section{Evaluation Methodology}
\label{sec:evaluation_methodology}

\begin{figure}[t]
\centering
\includegraphics[width=0.95\columnwidth]{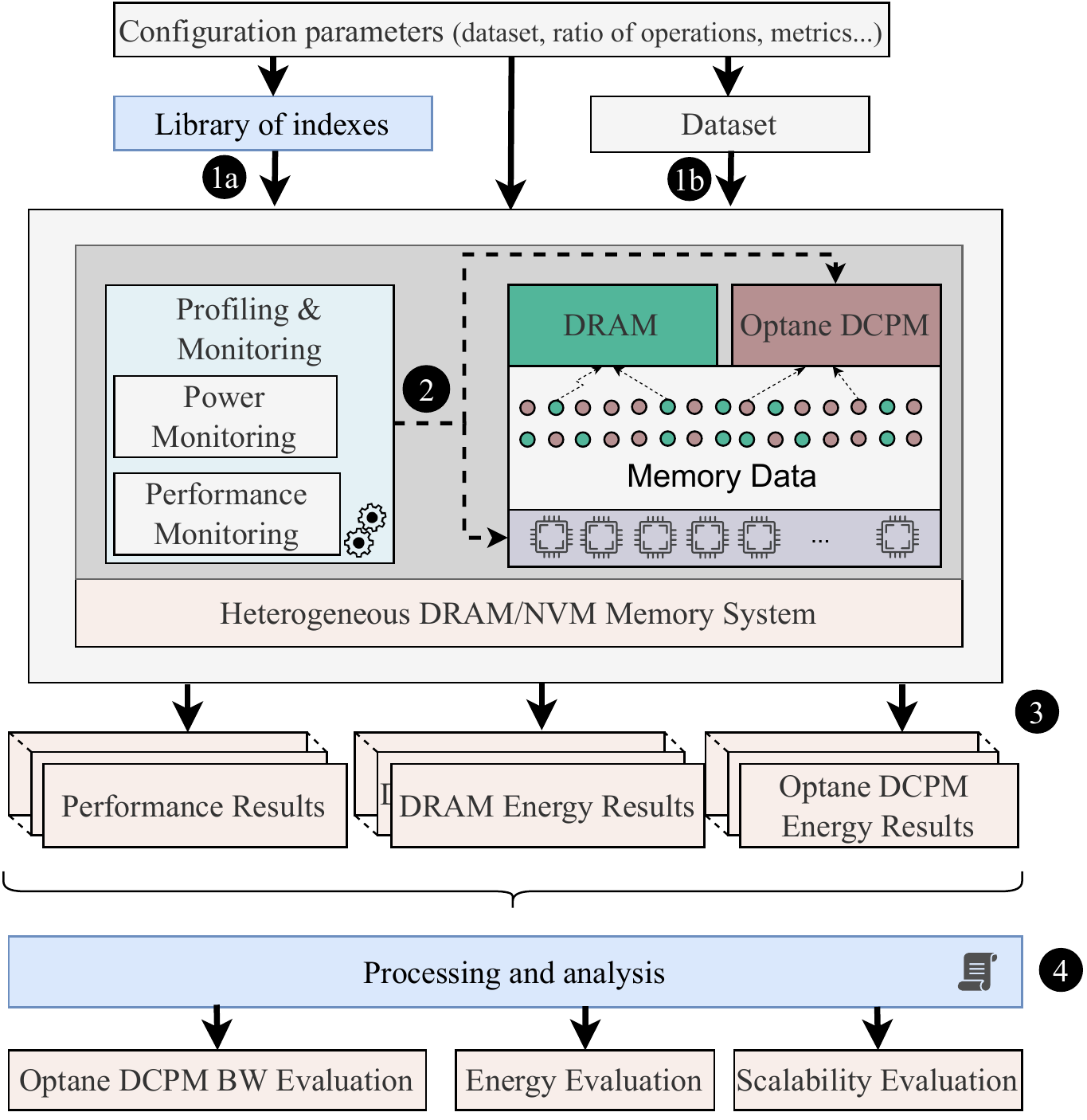}
\caption{Overview of the evaluation methodology}
\label{fig:methodology}
\end{figure}

In order to select representative B+ tree indexing implementations designed for persistent memories, we first identified their main design goals as described in the relevant literature. These goals are outlined below:
\begin{itemize}
    \item \textbf{Write operations minimization}. Write operations in the Optane DCPM are expensive in terms of latency. Therefore, indexes developed for persistent memories utilize an array of methods to minimize write operations like reduced logging when performing atomic operations, or avoiding the need for CPU cache flushes that are often needed to maintain consistency of the data on the persistent memory.
    \item \textbf{Efficient locking mechanisms}. In order to achieve high concurrency, read operations should be done in a consistent but efficient way across multiple threads to utilize the available bandwidth as much as possible. On the other hand, write operations should make use of as fine locks as possible, constraining the critical sections to necessary code only.
    \item \textbf{Efficient index traversal}. Several decisions play an important role in index traversal, like sorting, the type of tree, key splitting etc. Some indexes are implemented targeting increased data locality, while others store a part of the tree on DRAM for fast traversal.
\end{itemize}

These goals are materialized differently for each index implementation as they try to offer persistence and concurrency in an efficient way. Many different implementations have been proposed including trees, hash tables, tries, radix trees etc.~\cite{lee2019recipe}. In this work we focus on tree implementations, as they are the most commonly used in practice.

We design a systematic methodology to enable the effective execution of four widely used indexes in the literature, which is illustrated in Fig.~\ref{fig:methodology}.
The methodology consists of four distinct steps. 
The inputs of the methodology are a set of indexes~\circled{1a} and a dataset \circled{1b}, which are configured by a set of parameters, such as the workload type (ratio of operations, skewness, etc.), the selected indexes for evaluation and the metrics of interest (e.g. throughput, energy consumption, scalability). The library of indexes is based on the source codes provided by~\cite{lee2019recipe} and includes the B+ tree implementations listed below. These indexes were selected based on the  criterion of representing different ways of implementing the aforementioned design goals. 

\begin{itemize}
    \item \textbf{wBTree} is a type of main-memory B+ Tree, aiming to reduce the overhead caused by extensive NVM writes and CPU cache flush operations. To minimize the write operations wBTree uses write atomic B+ trees that can achieve node consistency either through atomic writes in the nodes or by redo-only logging, thus reducing the necessary write operations on Optane~\cite{chen2015persistent}. Furthermore, the keys are maintained in a sorted order throughout the execution and the nodes employ a small indirection slot array and/or a bitmap, to quickly check for a particular key's existence. The use of the indirection slot array has the added benefit of not requiring the movement of index entries, for most insertions and deletions, thus further reducing the amount of write operations. 
    \item \textbf{Masstree:} is a cache-efficient, highly concurrent trie-like concatenation of B+ tree nodes, and provides high performance even for long common key prefixes. We use the Masstree version converted for NVM by~\cite{lee2019recipe}. The Masstree combines a trie and B+ tree implementations to achieve higher performance for long keys. Particularly, long string keys are split into more than one node, so that searching is much faster, because of the faster comparisons between the common part of the keys. Masstree uses a versioning system for concurrency that improves locking performance. Furthermore, to reduce write operations, Masstree inserts new keys to leaf nodes by appending a new key-value pair to the node in an unsorted order. Internal nodes remain sorted for faster traversal.  
    \item \textbf{Fast\&Fair:}  is a persistent memory B+ tree that provides lock-free reads~\cite{lee2019recipe}. The reads detect and tolerate inconsistencies such as duplicated elements. By making read operations tolerate transient inconsistency, Fast\&Fair avoids expensive copy-on-write, logging, and even the necessity of read latches so that read transactions can be non-blocking~\cite{hwang2018endurable}. These characteristics of Fast\&Fair reduce write operations, while improving locking performance. 
    \item \textbf{FPTree:} is a B+ Tree alternative that implements selective persistence by requiring only leaf nodes to be stored persistently. Internal nodes can be rebuilt in case of power failure. Therefore, the FPTree places all internal nodes on the DRAM for fast traversing and consequently higher performance, while the leaf nodes are placed on a persistent memory. Furthermore, the leaf nodes implement \textit{fingerprinting}, where a one byte hash of each key contained within the node is placed at the first cacheline-sized section of the node~\cite{oukid2016fptree}. Fingerprinting improves lookup performance by speeding up the process of finding if a key is included in a particular node, without having to traverse the node itself. For concurrency, FPTree uses Hardware Transactional Memory (HTM)~\cite{yooTSX2013} and fine-grained locks for the internal and leaf nodes respectively. 
\end{itemize}

The \textit{Profiling \& Monitoring} component (\circled{2}) of the methodology is responsible for power and performance monitoring and it is based on Intel's Processor Counter Monitor (PCM)~\cite{pcm}, a tool that allows energy/power sampling over DIMMs through hardware counters~\cite{m2020rusty}. The sampling rate for the profiling is set to 0.01s, by default.
More specifically, we use the \textit{PCM-Power} component to monitor the energy consumed in the DRAM and Optane DCPM DIMMs. For each individual DIMM, the \textit{PCM-Power} reports the on-DIMM energy consumed for the period of time based on our defined sampling rate.
The memory accesses on the Optane DCPM are measured by the \textit{ipmctl} utility, which is the standard for configuring and managing Optane DCPM modules, by monitoring the corresponding read and write accesses of each DIMM, respectively~\cite{ipmctl}. 
Finally, to calculate the energy consumption of an SSD, we use profiling information about I/O reported by the Intel VTune Amplifier regarding the page read, write and flush operations and the power values from the manufacturers' data-sheets.

The measurements are grouped (\circled{3}) and processed (\circled{4}) in order to provide the throughput, energy consumption and scalability results for each evaluation experiment. The generation of the aforementioned results is performed automatically, based on a set of customizable Python and shell scripts. 
The source codes of the tool with usage instructions and the library of indexes are publicly available~\footnote{\url{{https://github.com/mkatsa/PENVMTool}}}.

\section{Evaluation}
\label{sec:evaluation}

\subsection{Experimental Setup}

Through the evaluation process we aim at the following: 
\begin{itemize}
    \item To evaluate the energy consumption behavior of the Optane DCPM for indexing data structures triggered by various types of workloads. 
    \item To study the performance and the energy efficiency of the Optane DPCM for a key-value store (LevelDB), using the corresponding deployment on an SSD as a baseline. 
\end{itemize}

\begin{table}[t]
\caption{Overview of experimental setup}
\resizebox{\columnwidth}{!}{
\begin{tabular}{|ll|}
\hline
\multicolumn{2}{|c|}{Experimental Setup}                                                                                                                                                                                    \\ \hline
\multicolumn{1}{|l|}{CPU, DRAM}                                                           & \begin{tabular}[c]{@{}l@{}}Xeon Gold 5218R CPU, 2x20 cores @2.10GHz,  \\ with 4x32GB DDR4 DIMMs \end{tabular} \\ \hline
\multicolumn{1}{|l|}{\begin{tabular}[c]{@{}l@{}}Optane DCPM\\ Configuration\end{tabular}} & \begin{tabular}[c]{@{}l@{}}2x256GB DIMMs, \\ AppDirect Mode, ext4-DAX\end{tabular}                                     \\ \hline
\multicolumn{1}{|l|}{SSD}                                                                 & Intel SSD S4500 Series, 480GB                                                                               \\ \hline
\multicolumn{1}{|l|}{Indexes}                                                             & wBTree, Masstree, Fast\&Fair, FPTree                                                                                            \\ \hline
\multicolumn{1}{|l|}{Key-value store}                                                             & LevelDB                                                                                            \\ \hline
\multicolumn{1}{|l|}{YCSB Workloads}                                                     & \begin{tabular}[c]{@{}l@{}}Balanced (50\% reads, 50\% writes), Write-heavy (99\% writes), \\ 
Read-heavy (100\% reads), Scan-heavy (99\% scan) \end{tabular} \\ \hline 
\multicolumn{1}{|l|}{Workload keys}                                                             & 8-bytes integers                                                                                            \\ \hline
\multicolumn{1}{|l|}{Operating system}                                                    & Ubuntu 20.04.2 LTS, kernel 5.4.0-121-generic                                                                                    \\ \hline
\end{tabular}
}
\label{tab:experimental_setup}
\end{table}

Table~\ref{tab:experimental_setup} shows an overview of the experimental setup. The experiments were conducted on a single node with a 2x20 core Intel Xeon Gold 5218R CPU @2.10GHz with 4x32GB DDR4 DIMMs and 2x256GB Optane DC NVDIMMs. Intel Optane DC is configured in \textit{App Direct} mode with ext4-DAX file system. We utilized the version 1.11 of PMDK~\cite{pmdk} and gcc-9.4. 
The monitoring and profiling tools were reported earlier, in Section \ref{sec:evaluation_methodology}.

The YCSB is used to trigger the indexes under different types of workload: balanced, write-heavy (which corresponds to insert operations), read-heavy (i.e. lookup operations) and scan-heavy (i.e. range-lookup operations)~\cite{cooper2010benchmarking}. Each workload consists of 64M operations with 8-byte random integer keys. 


\subsection{Baseline Performance-oriented Results}
 
First, we present a set of baseline performance results, which enable a comprehensive insight into the energy consumption evaluation, which follows.  

\begin{figure*}[t]
       \makebox[\linewidth]{ 
       \centering   
       \begin{subfigure}[]{0.255\textwidth}
                \includegraphics[width=\textwidth]{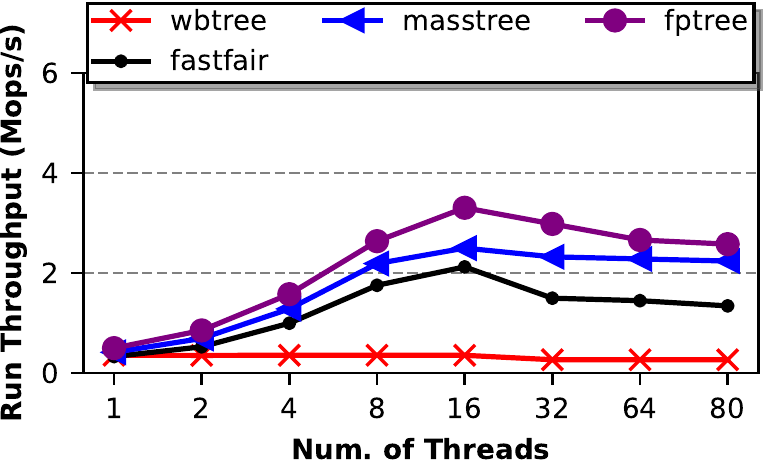}
                \caption{Balanced workload}
                \label{fig:throughput_ycsb_balanced}
        \end{subfigure}
         ~
         \begin{subfigure}[]{0.255\textwidth}
                \includegraphics[width=\textwidth]{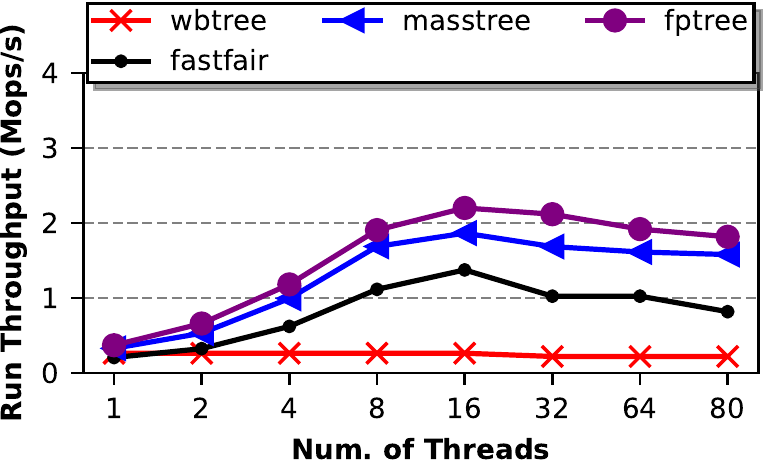}
                \caption{Write-heavy workload}
                \label{fig:throughput_ycsb_insert}
        \end{subfigure}
         ~
        \begin{subfigure}[]{0.25\textwidth}
                \includegraphics[width=\textwidth]{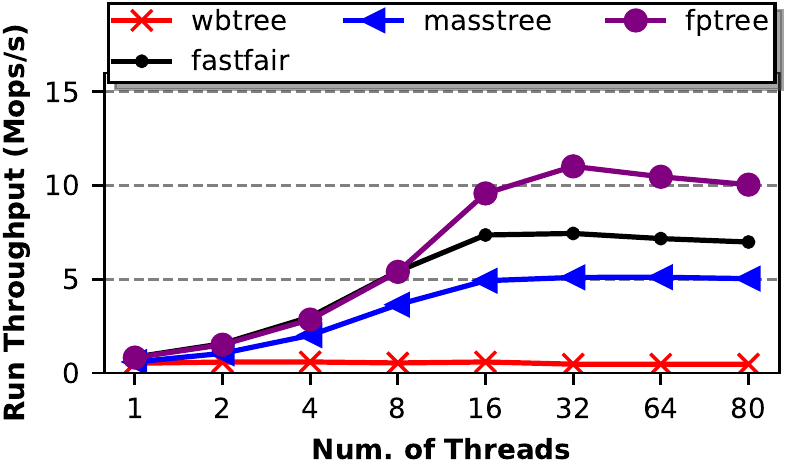}
                \caption{Read-heavy workload}
                \label{fig:throughput_ycsb_read}
        \end{subfigure}
         ~
        \begin{subfigure}[]{0.25\textwidth}
                \includegraphics[width=\textwidth]{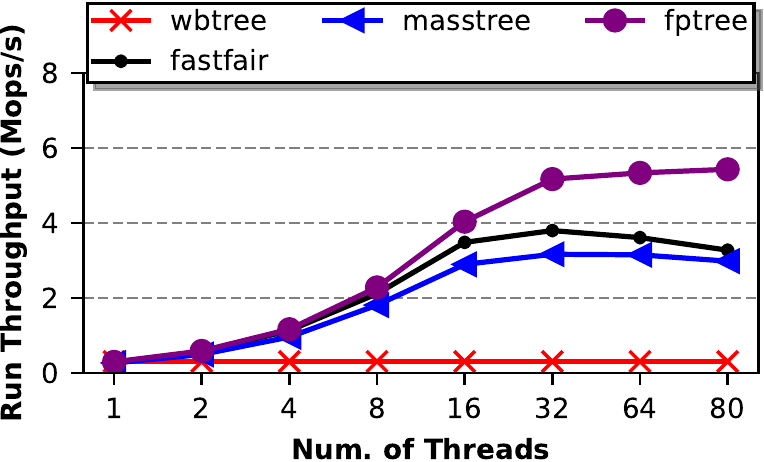}
                \caption{Scan-heavy workload}
                \label{fig:throughput_ycsb_scan}
        \end{subfigure}
    
        }
        \caption{Scalability evaluation on Intel Optane DCPM for different types of workloads.}
        \label{fig:throughput_ycsb}
\end{figure*}       

Fig.~\ref{fig:throughput_ycsb} shows the throughput on the Optane DCPM for a variety of YCSB workloads. It demonstrates the scalability of each index under different workloads. It can be noticed that the \textit{fptree} scales up to 32 threads for the read-heavy and scan-heavy workloads, by exceeding 10Mops/sec and 5Mops/sec, respectively. For the balanced and write-heavy workloads, the \textit{masstree}, \textit{fptree} and \textit{fastfair} provide maximum throughput up to 3.3Mops/sec and 2.3Mops/sec at 16 threads, respectively. The \textit{wbtree} does not scale, since the implementation used in this work is single-threaded and is demonstrated as a sequential baseline for the rest of the B+ tree implementations. 

\begin{figure}[t]
   \centering   
    \includegraphics[width=0.65\columnwidth]{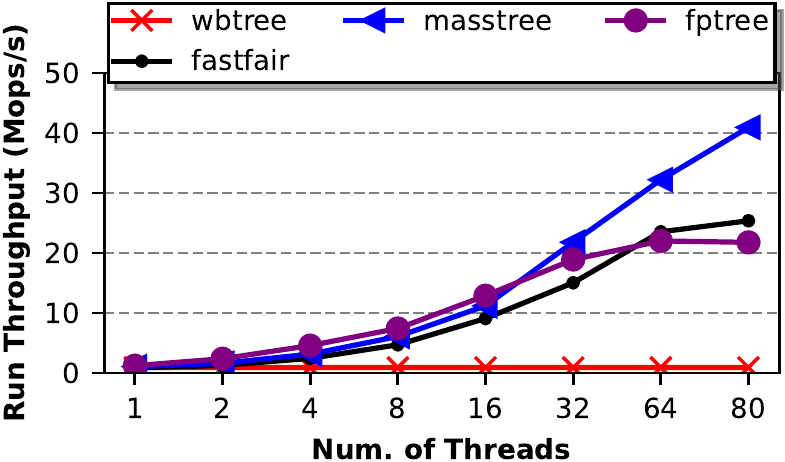}
    \label{fig:dram_run_randint_ycsb}
    \caption{Scalability evaluation on DRAM-only for balanced workload.} 
    \label{fig:dram_throughput_ycsb}
\end{figure}

For comparison, we evaluate the scalability of the indexes on a DRAM-only memory system. These baseline results are shown in Fig~\ref{fig:dram_throughput_ycsb}. All indexes, and especially the \textit{masstree}, scale well under the balanced workload up to 64 threads. Moreover we observe that the throughput achieved for \textit{masstree} index is greater than the \textit{fptree} on DRAM only, in contrast to the Optane DC execution, where \textit{fptree} outperforms \textit{masstree}. This is due to the fact that \textit{fptree} places all inner nodes on DRAM and only leaf nodes on Optane, thus performing better on higher number of threads especially on read-only workloads. Similar results were observed for the rest of the workload types. 

\begin{figure}[t]
\centering
\includegraphics[width=0.65\columnwidth]{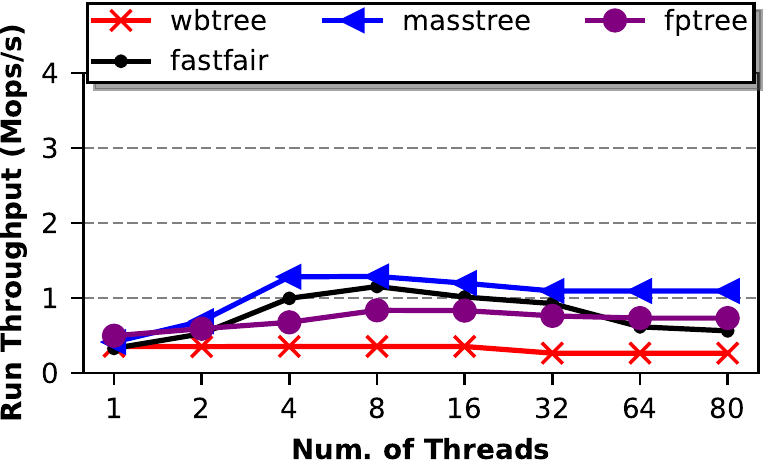}
\caption{Throughput for balanced workload using a single Optane DCPM DIMM.}
\label{fig:throughput_1DIMM}
\end{figure}

To investigate the impact of the Optane DCPM hardware configuration on the maximum throughput that can be reached, we evaluate the scalability performance of the indexes on a memory system with a single Optane DCPM DIMM, under balanced workload. By comparing the results of Fig.~\ref{fig:throughput_ycsb_balanced} (two DIMMs) with  Fig.~\ref{fig:throughput_1DIMM} (single DIMM), it is noticed that the use of more Optane DCPM DIMMs on the same socket enables higher parallelism, allowing improved scalability and higher throughput. Indeed, on a single DIMM, the maximum throughput is obtained at 4 threads and barely exceeds 1Mops/sec. 

 The above observations are inline with performance evaluation results in the existing literature. In particular, the scalability limitations of B+ tree indexes on Optane DCPM have been studied in recent works, relying on custom microbenchmarks~\cite{lersch2019evaluating}. Our findings, based on the YCSB, further confirm the observed scalability limitations. 
\subsection{Energy Consumption Results}

\begin{figure*}[t]
        \makebox[\linewidth]{ 
        \begin{subfigure}[]{0.25\textwidth}
                \includegraphics[width=\textwidth]{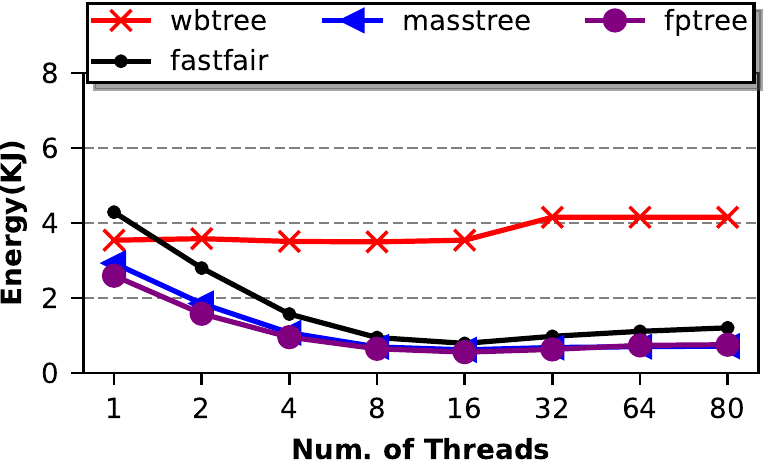}
                \caption{Balanced workload}
                \label{fig:energy_ycsb_balanced}
        \end{subfigure}
        ~
        \begin{subfigure}[]{0.25\textwidth}
                \includegraphics[width=\textwidth]{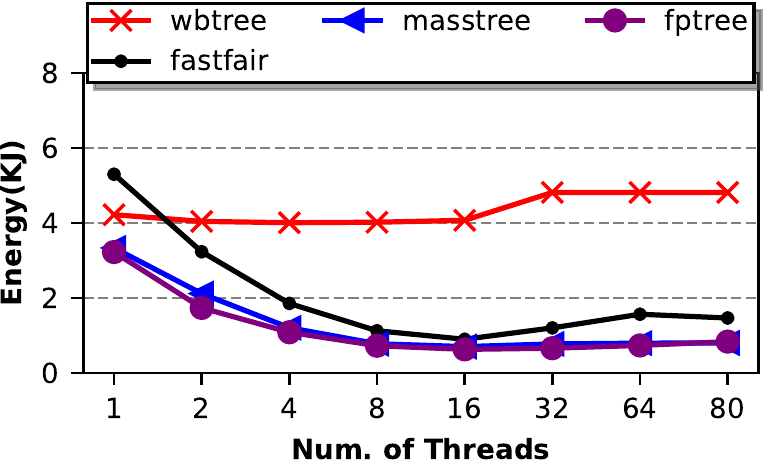}
                \caption{Write-heavy workload}
                \label{fig:energy_ycsb_insert}
        \end{subfigure}
        ~
        \begin{subfigure}[]{0.25\textwidth}
                \includegraphics[width=\textwidth]{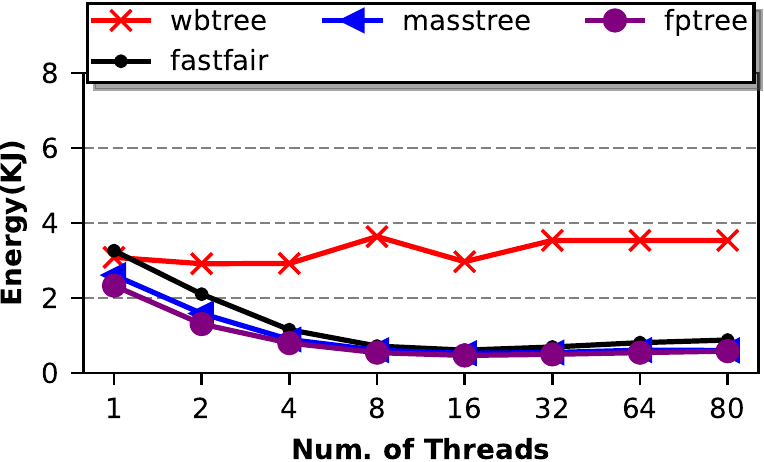}
                \caption{Read-heavy workload}
                \label{fig:energy_ycsb_read}
        \end{subfigure}
        ~
        \begin{subfigure}[]{0.25\textwidth}
                \includegraphics[width=\textwidth]{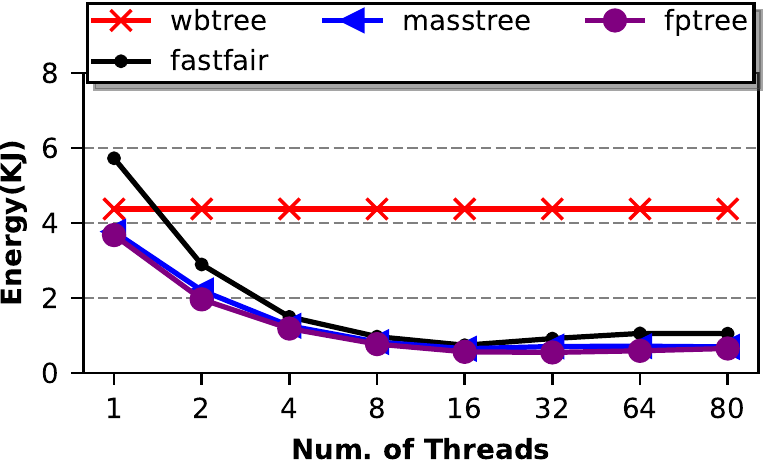}
                \caption{Scan-heavy workload}
                \label{fig:energy_ycsb_scan}
        \end{subfigure}
        }
        \caption{Energy consumption evaluation on Intel Optane DCPM for different types of workloads.}
        \label{fig:energy_ycsb}
\end{figure*}

Fig.~\ref{fig:energy_ycsb} shows the energy consumed in the Optane DCPM DIMMs for each index under different types of workload, as measured through the sensors equipped in the DIMMs. In general, the energy consumption is gradually reduced up to 8 or 16 threads, indicating a direct correlation between the throughput reached and the energy consumed. The energy of \textit{wbtree} index energy remains slightly constant. Since wbtree does not scale (Fig.~\ref{fig:throughput_ycsb}), this leads to utilization of the memory resources for higher amount of time, therefore leading to increased energy consumption.

\begin{table}[t]
\centering
\caption{Energy consumption (KJ) of Optane DCPM DIMMs for each workload type for 16 threads.}
\resizebox{\columnwidth}{!}{
\begin{tabular}{|c|c|c|c|c|}
\hline
         & Balanced & Write-heavy & Read-heavy & Scan \\ \hline
wbtree   & 3.539 & 4.069 & 2.966 & 4.372 \\ \hline
masstree & 0.615  & 0.701  & 0.524  & 0.645  \\ \hline
fptree   & 0.554  & 0.627  & 0.465  & 0.562  \\ \hline
fastfair & 0.787 & 0.896  & 0.620  & 0.738  \\ \hline
\end{tabular}
}
\label{tab:energy-per-workload}
\end{table}

Table \ref{tab:energy-per-workload} shows the energy consumption values for 16 threads. We notice that the energy consumption for different indexes under the same workload does not vary significantly. However, the Optane DCPM consumes less energy under the read-heavy workload compared to the rest of the workload types. For example, triggered by the read-heavy, the \textit{fptree} index yields 16\% lower energy consumption compared to the corresponding balanced workload, indicating the impact that the increased throughput under a read-heavy workload has on the energy (Fig.~\ref{fig:throughput_ycsb}).

\begin{figure}[t]
    \centering
    \includegraphics[width=0.65\columnwidth]{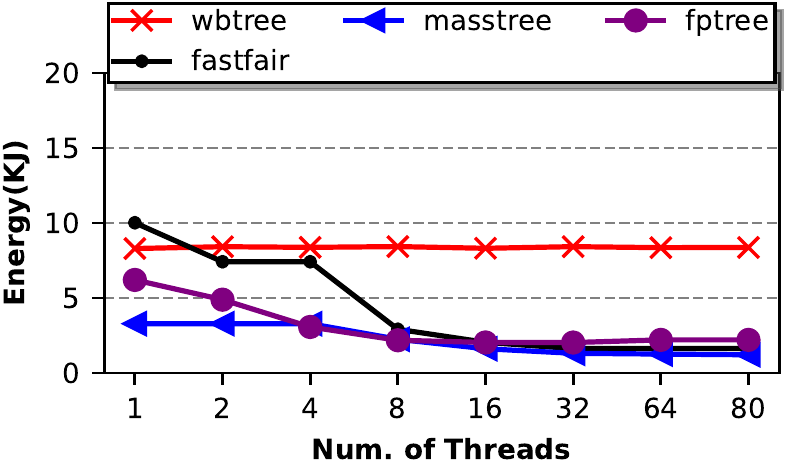}
    \caption{Energy consumption on DRAM-only for balanced workload.}
    \label{fig:dram_energy_ycsb}
\end{figure}

As a baseline for the Optane DCPM energy consumption, Fig.\ref{fig:dram_energy_ycsb} shows the energy for the balanced workload in a DRAM-only system (i.e. with Optane DCPM deactivated). Similarly to Fig.~\ref{fig:energy_ycsb}, the results correspond to the energy consumed on the DRAM DIMMs, only. The energy is significantly reduced after 4 or 8 threads, in accordance to the throughput increase demonstrated in Fig.~\ref{fig:dram_throughput_ycsb}. 
Additionally, by comparing the energy values of Fig.~\ref{fig:dram_energy_ycsb} with Fig.~\ref{fig:energy_ycsb_balanced}, we notice that the energy consumed in the Optane DCPM is about half the one in DRAM. As an example, the energy consumption when deploying the \textit{fptree} with 16 threads on Optane DCPM is 56.3\% lower compared to the corresponding DRAM experiment. 

\begin{figure}[t]
    \centering
    \includegraphics[width=0.65\columnwidth]{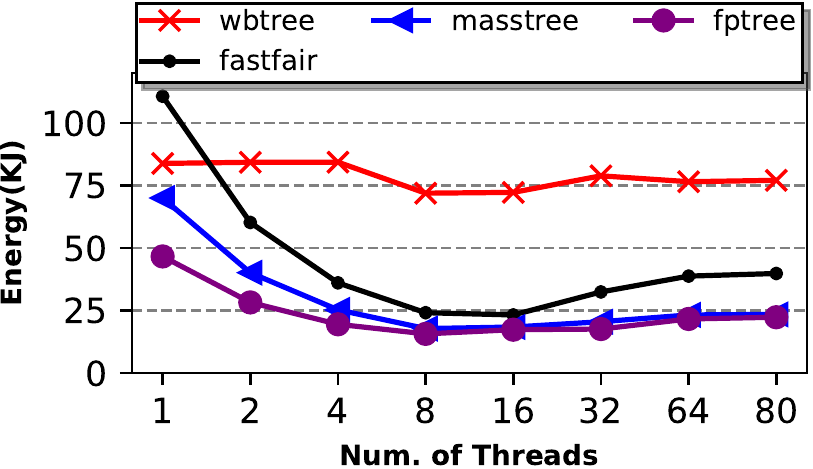}
    \caption{Energy consumption of CPUs and memory system.}
    \label{fig:optane-full-system}
\end{figure}

The total energy consumption, i.e. of both the CPUs and the memory system, is shown in Fig.~\ref{fig:optane-full-system}. The throughput still impacts the energy consumed, as it is minimized at 8-16 threads, where the maximum throughput is reached. However, after that point, the energy consumption slightly increases, as a result of the throughput decrease after 16 threads for most of the indexes, as depicted in Fig.~\ref{fig:throughput_ycsb}.

\begin{figure*}
\makebox[\textwidth]{ 
       \centering   
       \begin{subfigure}[]{0.75\textwidth}
                \includegraphics[width=\textwidth]{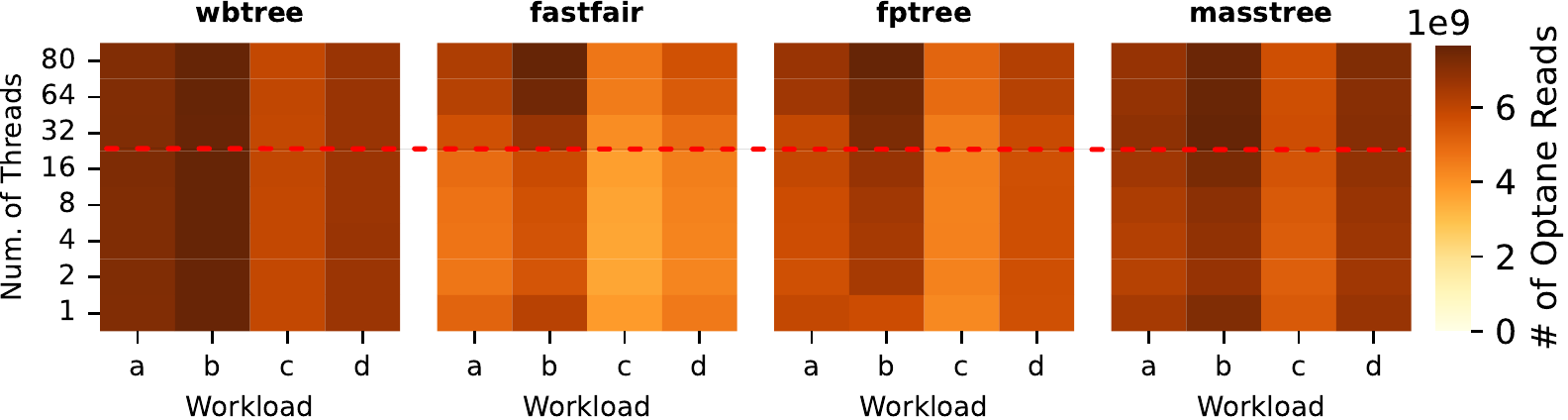}
                \caption{Intel Optane DCPM read accesses}
                \label{fig:read_accesses_ycsb}
        \end{subfigure}
        }
        
\makebox[\textwidth]{ 
        \begin{subfigure}[]{0.75\textwidth}
                \includegraphics[width=\textwidth]{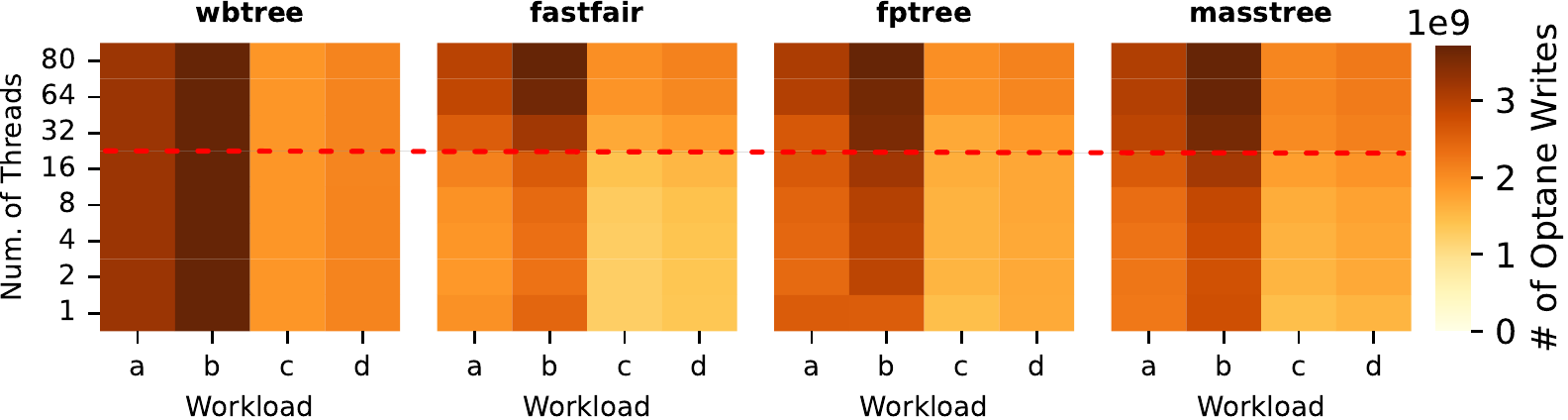}
                \caption{Intel Optane DCPM write accesses}
                \label{fig:write_accesses_ycsb}
        \end{subfigure}
}
        \caption{Number of Intel Optane DCPM read and write memory accesses. Workload a: balanced, b: write-heavy, c: read-heavy, d: scan-heavy.}
        \label{fig:memory_accesses}
\end{figure*}

To further investigate the energy consumption behavior of the Optane DCPM as the number of thread increases, we monitor the number of read and write memory accesses in the persistent memory for each index, for an increasing number of threads. The profiling results are shown in Fig.~\ref{fig:memory_accesses}. It is noticed that the number of read and write accesses beyond 16 threads increases significantly, due to the internal mechanisms of the indexes and the overhead of concurrency, without a positive impact on performance, as shown in Fig.\ref{fig:throughput_ycsb_balanced}. In particular, when the bandwidth is capped like in the case of 1 or 2 DIMMs of Optane DCPM, a significant amount of CPU time is spent by trying to acquire exclusive access to protected regions by each index, accessing internal index variables for locking and synchronization and thus increasing the overall access count. Therefore, using threads beyond that point has negative impact both on the throughput and the energy consumption. Additionally, the increase in the memory access count is expected to have impact the endurance of the persistent memory hardware~\cite{boukhobza2017emerging}. Interestingly, the writing hardware endurance of the Optane DCPM has not been evaluated in the existing literature, yet. 

\begin{figure}[t]
    \centering
    \includegraphics[width=0.65\columnwidth]{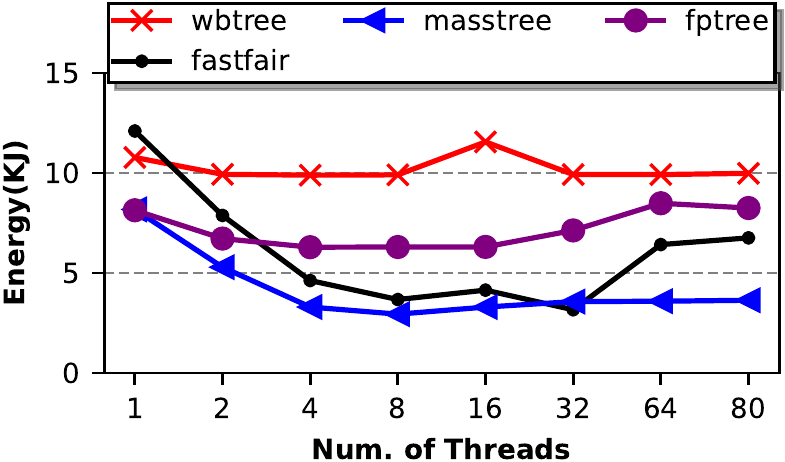}
    \caption{Energy consumption on a single Optane DCPM DIMM for balanced workload.}
    \label{fig:energy_ycsb_optane_1DIMM}
\end{figure}

Finally, to investigate the impact of the available persistent memory bandwidth on the energy consumption, Fig.~\ref{fig:energy_ycsb_optane_1DIMM} shows the energy for the balanced workload when integrating a single Optane DCPM DIMM. An interesting observation is that although when integrating 2 DIMMs the throughput increases almost 2x (as shown by Fig.\ref{fig:throughput_ycsb_balanced} and Fig.\ref{fig:throughput_1DIMM}), the energy is reduced by 5x (e.g. for the \textit{masstree} as shown by comparing the results of Fig.\ref{fig:energy_ycsb_balanced} and Fig.~\ref{fig:energy_ycsb_optane_1DIMM}). Therefore, it is reasonable to expect that increasing the Optane DCPM bandwidth by integrating even more DIMMs will further decrease the energy consumption and increase the throughput. 

\subsection{LevelDB key-value store Performance-Energy Exploration}

In this section, we perform a combined performance-energy consumption analysis of the Optane DCPM using the LevelDB key-value store\footnote{LevelDB: https://github.com/google/leveldb}. 

The LevelDB is a fast key-value storage library developed by Google, where keys and values are arbitrary byte arrays, allowing any type of data to be sorted by a key. LevelDB is not an SQL database and it does not provide an interface for SQL queries making it more inline with the other indexes described in this work. The only difference is that the LevelDB keeps data stored on disk, while maintaining a configurable buffer in the memory for performance. The inherent support for data persistence on HDD/SSD and data caching in DRAM make it a suitable candidate for demonstrating the benefits of storing the whole database in an Optane DCPM instead of an SSD. The LevelDB uses a log-structured merge-tree (also known as LSM tree) as a data structure for storing data. The LSM tree comprises two parts: one tree in the memory serving as a buffer and another on the disk. The advantage and particular characteristic of the LSM trees is that each part can be configured differently for the platform they reside.

\begin{figure}[t]
       \begin{minipage}[]{\linewidth}
       \centering   
        \begin{subfigure}[]{0.4\columnwidth}
                \includegraphics[width=\textwidth]{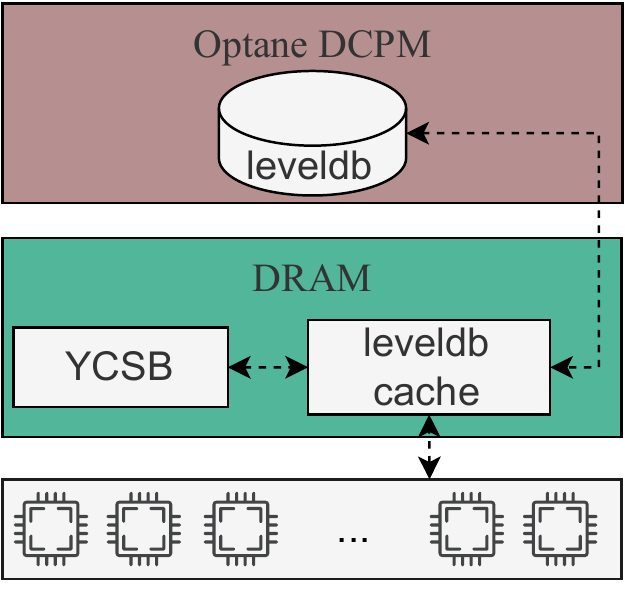}
                \caption{DRAM/SSD configuration.}
                \label{fig:leveldb_optane}
        \end{subfigure}
        ~
        \begin{subfigure}[]{0.4\linewidth}
                \includegraphics[width=\textwidth]{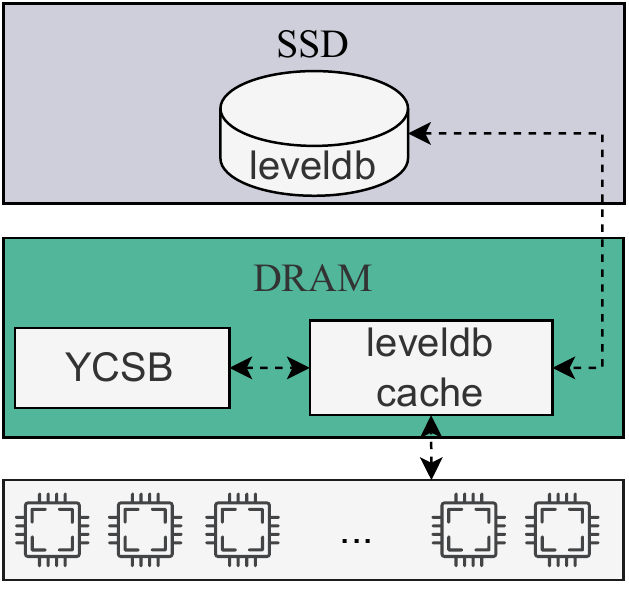}
                \caption{DRAM/Optane DCPM configuration.}
                \label{fig:leveldb_disk}
        \end{subfigure}
        \caption{System configurations for the LevelDB experiments: DRAM/SSD and DRAM/Optane DCPM.}
        \label{fig:leveldb-memory-setup}
        \end{minipage} 
\end{figure}

After integrating the YCSB in the LevelDB, we evaluate the two memory hierarchy configurations depicted on Fig.~\ref{fig:leveldb-memory-setup}, in terms of throughput and energy consumption, under different workloads. In the first experiment, the LevelDB is stored on the SSD, while in the second it is allocated in the Optane DCPM. The tools employed for profiling and monitoring are described earlier, in Section~\ref{sec:evaluation_methodology}. For brevity, we present results for the balanced and read-heavy workloads only, but the observations apply to the write-heavy and scan workloads, as well. 

\begin{figure}[t]
       \centering   
        \begin{subfigure}[]{0.48\columnwidth}
        \includegraphics[width=\columnwidth]{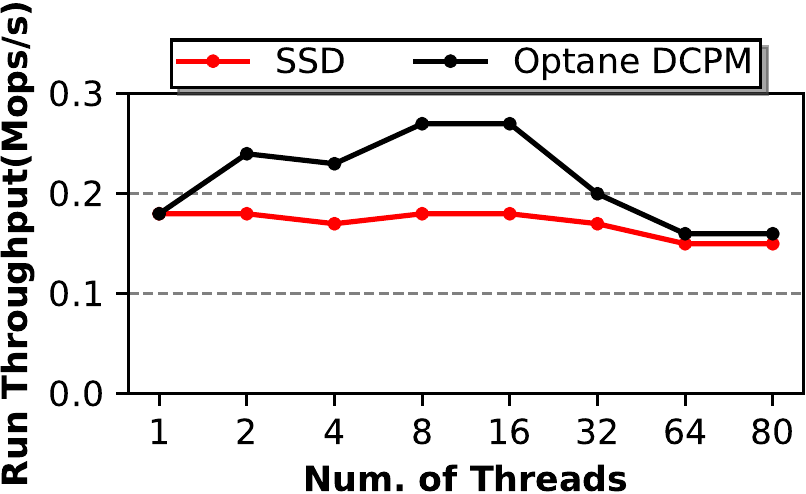}
        \caption{Balanced workload}
         \label{fig:leveldb-throughput-balanced}
        \end{subfigure}
        ~
       \begin{subfigure}[]{0.48\columnwidth}
        \includegraphics[width=\columnwidth]{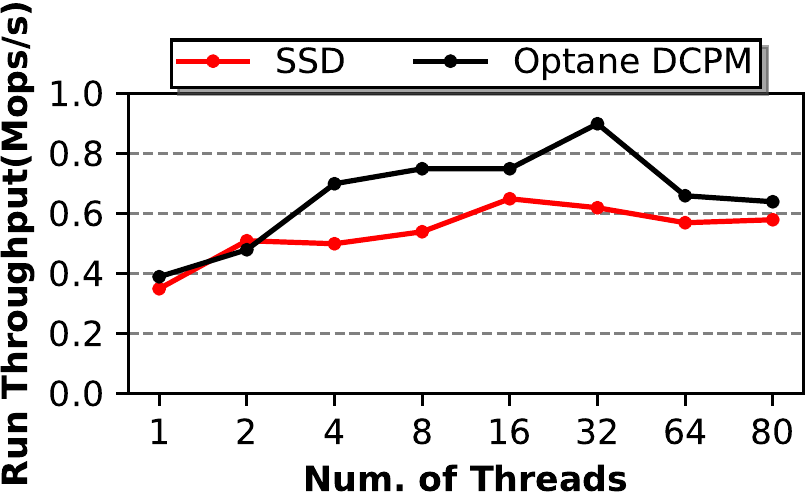}
        \caption{Read-heavy workload}
     \label{fig:leveldb-throughput-read-heavy}
        \end{subfigure}
        \caption{LevelDB throughput for different memory configurations under balanced and read-heavy workload.}
        \label{fig:leveldb-throughput}
\end{figure}

The throughput results are presented in Fig.~\ref{fig:leveldb-throughput-balanced} and Fig.~\ref{fig:leveldb-throughput-read-heavy}, for balanced and read-heavy workloads, respectively. We notice the following: 
\begin{itemize}
    \item The maximum throughput for the Optane DCPM is reached at 16 threads for the balanced workload and at 32 for the read-heavy. This result is inline with the observations based on the indexing data structure experiments on Optane DCPM (Fig.~\ref{fig:throughput_ycsb_balanced} and Fig.~\ref{fig:throughput_ycsb_read}). For the LevelDB, increasing the number of threads beyond these points, results in performance degradation on the Optane DCPM, as noticed in most of the corresponding experiments using B+ tree indexes (Fig.~\ref{fig:throughput_ycsb}).  
    \item Increasing the number of threads beyond the point of the maximum throughput, results in an increasing memory access count on the Optane DCPM. Indeed, for the balanced workload, the number of accesses increases by 27.4\% for 32 threads and  by 39.1\% for 64, compared to the number of accesses on 16 threads. Similarly, for the read-heavy workload it increases by 13.6\% for 64 threads compared to 32. These results correspond to the ones demonstrated in Fig.~\ref{fig:memory_accesses} for the indexing data structures experiments. The increase in the memory access count is attributed to the overhead of the synchronization mechanisms and is expected to have negative impact on the Optane DCPM hardware endurance. 
    \item Finally, Fig.~\ref{fig:leveldb-throughput-balanced} and Fig.~\ref{fig:leveldb-throughput-read-heavy} demonstrate the performance gains which can be obtained for the LevelDB when stored in an Optane DCPM instead of an SSD. Performance gains in terms of throughput reach 37.3\% for the balanced workload (at 16 threads) and 34.6\% for the read-heavy at 32 threads. 
\end{itemize}

The energy consumption results are presented in Fig.\ref{fig:leveldb-energy-balanced} and Fig.\ref{fig:leveldb-energy-read-heavy} for the balanced and the read-heavy workloads, respectively. Detailed measurements on (i) CPUs, DRAM DIMMs and SSD, (ii) Optane DIMMs only, (iii) CPUs, DRAM and Optane DIMMs are presented. We observe the following: 
\begin{itemize}
\item As the number of threads increases, the total energy consumption for the balanced workload for the Optane DCPM experiment (i.e. CPU, DRAM, Optane DCPM) slightly drops reaching a minimum value of 2.6KJ at 8 threads, which is 16\% lower than the energy for a single thread. The energy consumption on the Optane DCPM DIMMs is also in direct correlation with the throughput (Fig.\ref{fig:leveldb-throughput-balanced}). After that point, the energy increases following the decrease in throughput. In contrast with the Optane DCPM experiments, the energy consumption when the LevelDB is stored in the SSD is relatively independent of the number of threads for the balanced workload, slightly decreasing at 16 threads, by 2.1\% compared to the single thread experiment. 
\item For the read-heavy workload, the total energy consumption for the Optane DCPM experiment (i.e. CPU, DRAM, Optane DCPM) drops to 0.7KJ at 32 threads, where the throughput is maximized. Compared to the single thread experiment, the energy is decreased by 31\%. The energy consumed by the Optane DIMMs only, is also minimized at 32 threads. Additionally, the energy when storing the LevelDB to the SSD drops up to 32 threads and increases after that point. 
Another interesting observation is the fact that the energy consumed by the Optane DCPM is about the same between the experiments of different workload types (i.e. about 0.5KJ for both the balanced and read-heavy workloads). For the read-heavy workload the energy consumed by the CPU and DRAM DIMMs is between 0.4-0.6KJ, depending on the number of threads, as shown in Fig.~\ref{fig:leveldb-energy-read-heavy}. However, for the balanced workload, a much larger portion of the total energy is consumed by the CPU and the DRAM (about 2.5KJ, which is almost 6x higher than the energy consumed by the Optane DCPM DIMMs only, as shown in Fig.~\ref{fig:leveldb-energy-balanced}).  
\item Significant energy gains can be obtained by storing the LevelDB to an  Optane DCPM, instead of an SSD. For example, for the read-heavy workload, the energy consumption is reduced up to 71.2\%. This demonstrates the efficiency of Optane DCPM as a storage medium, it terms of energy. 
\end{itemize}


\subsection{Discussion}

In this subsection we highlight the main key findings based on the analysis of the experimental results and we indicate some major open research directions. 

The Optane DCPM is a relevant persistent memory architecture for data intensive HPC applications, providing advantages in terms of both throughput and energy consumption compared to the typical SSD storage. However, the bandwidth limitations of the Optane DCPM need to be taken into consideration. The evaluation results indicate that \textbf{fine-tuning the deployment of the indexing data structures on an heterogeneous memory system} is needed, in order to provide significant gains in terms of combined energy consumption and performance. 

\textbf{The hardware configuration is critical for the scalability and the energy consumption of applications deployed on the Optane DCPM}. Based on the experiments of Fig.~\ref{fig:throughput_1DIMM} and Fig.~\ref{fig:throughput_ycsb_balanced}, we can argue that by integrating more DIMMs into the memory system, higher throughput will be reached due to improved parallelism. It is reasonable to expect that improvements in throughput will result in lower energy consumption. However, it will be interesting to investigate the existence of a saturation point, after which integrating more DIMMs does not contribute to further energy efficiency.

\begin{figure}[t]
       \centering   
        \begin{subfigure}[]{0.48\linewidth}
        \includegraphics[width=\linewidth]{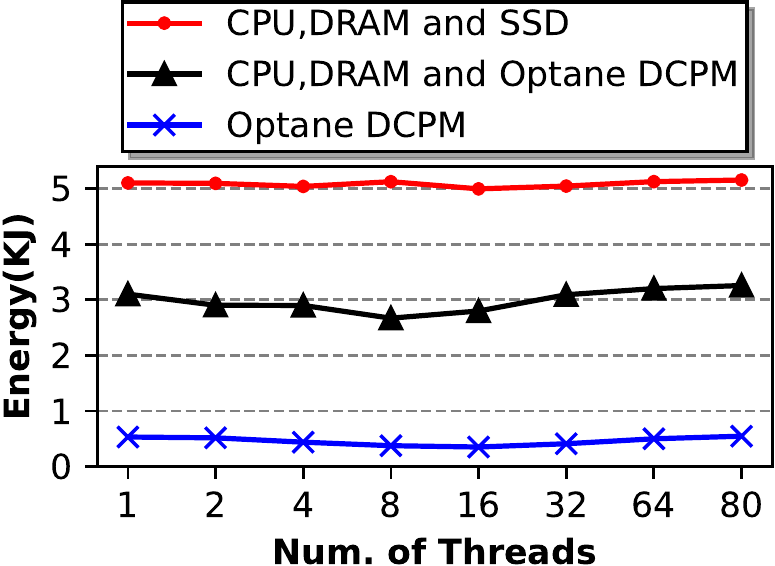}
        \caption{Balanced workload}
         \label{fig:leveldb-energy-balanced}
        \end{subfigure}
        ~
       \begin{subfigure}[]{0.48\linewidth}
        \includegraphics[width=\linewidth]{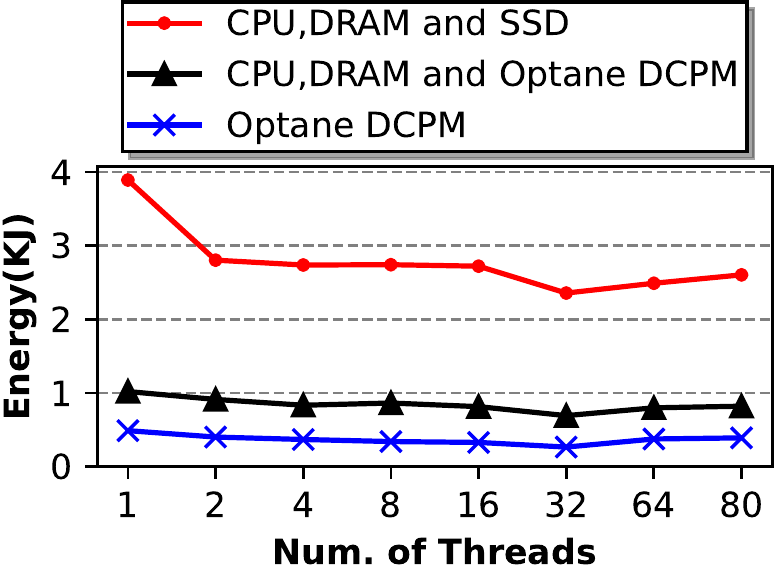}
        \caption{Read-heavy workload}
     \label{fig:leveldb-energy-read-heavy}
        \end{subfigure}
        \caption{LevelDB energy consumption for different memory configurations under balanced and read-heavy workload.}
        \label{fig:leveldb-energy}
\end{figure}

\textbf{The write indexing operations are costly in terms of energy consumption} on the Optane DCPM compared to the read-only (e.g. lookup operations). However, the increased amount of energy is not consumed by the Optane DIMMs, as demonstrated in the LevelDB experiments of Fig.~\ref{fig:leveldb-energy} (but also in Fig.~\ref{fig:energy_ycsb}). Instead, it is consumed by the CPUs, due to the synchronization and consistency overhead, requiring cache flushes and similar instructions to maintain the data coherence in the persistent device. This energy consumption overhead can be observed both in the B+ tree and LevelDB experiments. Hardware support for data consistency targeting NVMs is expected to benefit the energy efficiency of the whole system significantly. 

Although the analysis of the experimental results was based on experiments using 8-byte integer keys, \textbf{similar observations can be obtained by the use of other types of keys}. As an indicative example, for 24-byte string keys on Optane DCPM, using the same experimental setup, we noticed the energy consumed by the Optane DCPM DIMMs to follow a similar pattern with the corresponding integer-key experiments. Particularly, it is reduce up to 8 and 16 threads for the balanced and read-heavy workloads respectively, and reaches a plateau after that point. 

During the profiling analysis of Section~\ref{sec:evaluation}, we have evaluated the scalability of the dynamic allocations using the \textit{libvmmalloc} library of the PMDK~\cite{pmdk}. We noticed that the \textbf{\textit{libvmmalloc} does not scale with the number of threads}. This is inline with observations in the existing literature~\cite{lersch2019evaluating}. This limitation of the \textit{libvmmalloc} will affect the scalability and the energy consumption of the indexing data structures on the Optane DCPM, as well as the performance of dynamic applications which make heavy use of the heap memory and are deployed to Optane DCPM through the PMDK libraries. Custom and scalable dynamic memory allocators for persistent memories are expected to enable higher performance and lower energy consumption for such applications. 

\section{Conclusion}
\label{sec:conclusion}

This work provides new insights about the Optane DCPM as a persistent storage device. Based on a comprehensive evaluation of indexing data structures triggered by various types of workloads, we investigate the energy consumption characteristics of the Optane DCPM. We show that significant performance and energy consumption gains can be obtained by the effective use of Optane DCPM integrated in a memory system, compared to a typical SSD. However, proper configuration is required, considering the bandwidth limitations of the Optane DCPM, in order to benefit from the advantages it offers. The outcomes of this work can be exploited by developers who target the Optane DCPM as a persistent storage medium for HPC applications.


\bibliographystyle{./bibliography/IEEEtran}
\bibliography{./bibliography/IEEEabrv,./bibliography/mybib}

\end{document}